\documentstyle[preprint,aps,tighten]{revtex}
\begin{document}
\preprint{cond-mat/9403055}
\draft
\title{Scaling theory of conduction\\
through a normal--superconductor microbridge}
\author{C. W. J. Beenakker, B. Rejaei, and J. A. Melsen}
\address{Instituut-Lorentz, University of Leiden\\
P.O. Box 9506, 2300 RA Leiden, The Netherlands}
\maketitle
\begin{abstract}
The length dependence is computed of the resistance of a disordered
normal-metal wire attached to a superconductor. The scaling of the transmission
eigenvalue distribution with length is obtained exactly in the metallic limit,
by a transformation onto the isobaric flow of a two-dimensional ideal fluid.
The resistance has a {\em minimum\/} for lengths near $l/\Gamma$, with $l$ the
mean free path and $\Gamma$ the transmittance of the superconductor interface.
\end{abstract}
\pacs{PACS numbers: 74.80.Fp, 72.15.Rn, 73.40.Gk, 74.50.+r}
\narrowtext

The resistance of a disordered wire increases with its length. In the metallic
regime (length $L$ much less than the localization length) the increase is
linear, up to relatively small quantum corrections. The linear scaling seems
obviously true regardless of the nature of the contacts to the wire, which
would just contribute an additive, $L$-independent contact resistance according
to Ohm's law. This is correct if the contacts are in the normal state. The
purpose of this paper is to demonstrate and explain the complete breakdown of
the linear scaling if one of the contacts is in the superconducting state. The
resistance $R_{\rm NS}(L)$ of a disordered normal-metal (N) wire (with mean
free path $l$) connected to a superconductor (S) by a tunnel barrier (with
transmission probability $\Gamma$) has a {\em minimum\/} when $L\simeq
l/\Gamma$.

The existence of a resistance minimum, and its destruction by a voltage or
magnetic field, was first proposed by Van Wees et al.\ \cite{Wee92}, to explain
the sharp dip in the differential resistance discovered by Kastalsky et al.\
\cite{Kas91}. A similar anomaly has since been observed in a variety of
semiconductor--superconductor junctions \cite{exp}, and also in computer
simulations \cite{Tak93,Mar93}. A non-monotonous $L$-dependence of $R_{\rm NS}$
is implicit in the work of Volkov, Zaitsev, and Klapwijk \cite{Vol93}, while
Hekking and Nazarov \cite{Hek93} obtained a monotonous $R_{\rm NS}(L)$ for a
high tunnel barrier. Very recently, Nazarov \cite{Naz93}, using a Green's
function technique, has confirmed the results of Ref.\ \cite{Vol93}, in a
highly interesting paper which has some overlap with the present work.

Our analysis builds on two theoretical results. The first result we use is a
Landauer-type formula \cite{Bee92} for the conductance $G_{\rm NS}\equiv
1/R_{\rm NS}$,
\begin{equation}
G_{\rm NS}=\frac{4e^{2}}{h}\sum_{n=1}^{N}
\left(\frac{T_{n}}{2-T_{n}}\right)^{2}.\label{GNS}
\end{equation}
The numbers $T_{n}\in[0,1]$ are the eigenvalues of the matrix product
$tt^{\dagger}$, with $t$ the $N\times N$ transmission matrix of the disordered
normal wire plus tunnel barrier ($N$ being the number of transverse modes at
the Fermi energy $E_{\rm F}$). Eq.\ (\ref{GNS}) holds in the zero-temperature,
zero-voltage, and zero-magnetic field limit. Terms of order $(\Delta/E_{\rm
F})^{2}$ are neglected (with $\Delta$ the superconducting energy gap), as well
as contributions from disorder in the superconductor. [Disorder in the
superconductor will increase the effective length $L$ of the disordered region
by an amount of the order of the superconducting coherence length.]

The second result we use is a scaling equation \cite{Mel89},
\begin{eqnarray}
\frac{\partial}{\partial s}\rho(\lambda,s)&=&-\frac{2}{N}
\frac{\partial}{\partial\lambda}\lambda(1+\lambda)\rho(\lambda,s)
\frac{\partial}{\partial\lambda}
\int_{0}^{\infty}\!\!d\lambda'\,
\rho(\lambda',s)\nonumber\\
&&\hspace{3.5cm}\mbox{}\times\ln|\lambda-\lambda'|,\label{scaling}
\end{eqnarray}
where $s\equiv L/l$. We have made the conventional change of variables from
$T_{n}$ to $\lambda_{n}\equiv(1-T_{n})/T_{n}$, with $\lambda_{n}\in[0,\infty)$.
The density $\rho$ of the $\lambda$'s is defined by
$\rho(\lambda,L)=\langle\sum_{n}\delta(\lambda-\lambda_{n})$,
with $\langle\cdots\rangle$ the ensemble average. Eq.\ (\ref{scaling})
describes the scaling of the eigenvalue density with the length $L$ of the
disordered normal region. In this formulation the tunnel barrier at the NS
interface appears as an initial condition,
\begin{equation}
\rho(\lambda,0)=
N\delta\biglb(\lambda-(1-\Gamma)/\Gamma\bigrb),\label{rhoinitial}
\end{equation}
where $\Gamma$ is the transmission probability of the barrier. (For simplicity
we assume a mode-independent transmission probability, i.e.\ all $T_{n}$'s
equal to $\Gamma$ when $L=0$.) Once $\rho$ is known, one can compute the
ensemble averaged conductance $\langle G_{\rm NS}\rangle$ from Eq.\
(\ref{GNS}),
\begin{equation}
\langle G_{\rm NS}\rangle=\frac{4e^{2}}{h}
\int_{0}^{\infty}\!\!d\lambda\,\rho(\lambda,s)(1+2\lambda)^{-2}.
\label{GNSaverage}
\end{equation}

The non-linear diffusion equation (\ref{scaling}) was derived by Mello and
Pichard \cite{Mel89} from a Fokker-Planck equation \cite{Dor82,Sto91} for the
joint distribution function of all $N$ eigenvalues, by integrating out $N-1$
eigenvalues and taking the large-$N$ limit. This limit restricts its validity
to the metallic regime ($N\gg L/l$), and is sufficient to determine the leading
order contribution to the average conductance, which is ${\cal O}(N)$. The
weak-localization correction, which is ${\cal O}(1)$, is neglected
\cite{Note2}. These ${\cal O}(1)$ corrections depend on the magnetic field $B$,
whereas the ${\cal O}(N)$ contributions to $\rho$ described by Eq.\
(\ref{scaling}) are insensitive to $B$. However, the relationship
(\ref{GNSaverage}) between $G_{\rm NS}$ and $\rho$ holds only for $B=0$. (For
$B\neq 0$, $G_{\rm NS}$ depends not only on the eigenvalues of $tt^{\dagger}$,
but also on the eigenvectors \cite{Bee92}.) A priori, Eq.\ (\ref{scaling})
holds only for a wire geometry (length $L$ much greater than width $W$),
because the Fokker-Planck equation from which it is derived \cite{Sto91}
requires $L\gg W$. Numerical simulations \cite{Sle93} indicate that the
geometry dependence only appears in the ${\cal O}(1)$ corrections, and that the
${\cal O}(N)$ contributions are essentially the same for a wire, square, or
cube.

Our method of solution is a variation on Carleman's method \cite{Car22}. We
introduce an auxiliary function
\begin{equation}
F(z,s)=\int_{0}^{\infty}\!\!d\lambda'\,
\frac{\rho(\lambda',s)}{z-\lambda'},\label{Fdef}
\end{equation}
which is analytic in the complex $z$-plane cut by the positive real axis.
Furthermore,
\begin{equation}
\lim_{|z|\rightarrow\infty}F(z,s)=N/z.\label{Flimit}
\end{equation}
The function $F$ has a discontinuity for $z=\lambda\pm{\rm i}\epsilon$ (with
$\lambda>0$ and $\epsilon$ a positive infinitesimal). The limiting values
$F_{\pm}(\lambda,s)\equiv F(\lambda\pm{\rm i}\epsilon,s)$ are
\begin{equation}
F_{\pm}=\pm\frac{\pi}{\rm i}\rho(\lambda,s)+\frac{\partial}{\partial\lambda}
\int_{0}^{\infty}\!\!d\lambda'\,\rho(\lambda',s)\ln|\lambda-\lambda'|.
\label{Fdisc}
\end{equation}
Combination of Eqs.\ (\ref{scaling}) and (\ref{Fdisc}) gives
\begin{equation}
N\frac{\partial}{\partial s}(F_{+}-F_{-})=-\frac{\partial}
{\partial\lambda}\lambda(1+\lambda) (F_{+}^{2}-F_{-}^{2}),\label{scaling3}
\end{equation}
which implies that the function
\begin{equation}
{\cal F}(z,s)=N\frac{\partial}{\partial s}F(z,s)+\frac{\partial} {\partial
z}z(1+z)F^{2}(z,s)\label{calFdef}
\end{equation}
is analytic in the whole complex plane, including the real axis. Moreover,
${\cal F}\rightarrow 0$ for $|z|\rightarrow\infty$, in view of Eq.\
(\ref{Flimit}). We conclude that ${\cal F}\equiv 0$, since the only analytic
function which vanishes at infinity is identically zero.

It is convenient to make the mapping $z=\sinh^{2}\zeta$ of the $z$-plane onto
the strip ${\cal S}$ in the $\zeta$-plane between the lines $y=0$ and
$y=-\pi/2$, where $\zeta=x+{\rm i}y$. The mapping is conformal if we cut the
$z$-plane by the two halflines $\lambda>0$ and $\lambda<-1$ on the real axis.
On ${\cal S}$ we define the auxiliary function $U=U_{x}+{\rm i}U_{y}$ by
\begin{equation}
U(\zeta,s)\equiv\frac{F}{2N}\,\frac{dz}{d\zeta}
=\frac{\sinh 2\zeta}{2N}
\int_{0}^{\infty}\!\!d\lambda'\,\frac{\rho(\lambda',s)}
{\sinh^{2}\zeta-\lambda'}.\label{Udef}
\end{equation}
The equation ${\cal F}\equiv 0$ now takes the form
\begin{equation}
\frac{\partial}{\partial s}U(\zeta,s)+U(\zeta,s)\frac{\partial} {\partial
\zeta}U(\zeta,s)=0,\label{Euler}
\end{equation}
which we recognize as {\em Euler's equation\/} of hydrodynamics:
$(U_{x},U_{y})$ is the velocity field in the $(x,y)$ plane of a two-dimensional
ideal fluid at constant pressure.

Euler's equation is easily solved. For initial condition
$U(\zeta,0)=U_{0}(\zeta)$ the solution to Eq.\ (\ref{Euler}) is
\begin{equation}
U(\zeta,s)=U_{0}\biglb(\zeta-sU(\zeta,s)\bigrb).\label{Eulersol}
\end{equation}
To restrict the flow to ${\cal S}$ we demand that both $\zeta$ and
$\zeta-sU(\zeta,s)$ lie in ${\cal S}$. From $U$ we obtain the eigenvalue
density, first in the $x$-variables
\begin{equation}
\rho(x,s)=(2N/\pi)U_{y}(x-{\rm i}\epsilon,s),\label{rhoxU}
\end{equation}
and then in the $\lambda$-variables ($\lambda=\sinh^{2}x$),
\begin{equation}
\rho(\lambda,s)\equiv\rho(x,s)|dx/d\lambda|= \rho(x,s)|\sinh
2x|^{-1}.\label{rholambdaU}
\end{equation}
Eqs.\ (\ref{Eulersol})--(\ref{rholambdaU}) represent the {\em exact\/} solution
of the non-linear diffusion equation (\ref{scaling}), for {\em arbitrary\/}
initial conditions.

The initial condition (\ref{rhoinitial}) corresponds to
\begin{equation}
U_{0}(\zeta)=\case{1}{2}\sinh 2\zeta\,(\cosh^{2}\zeta
-\Gamma^{-1})^{-1}.\label{U0Gamma}
\end{equation}
The solution of the implicit equation (\ref{Eulersol}) is plotted in Fig.\
\ref{rhoplot} (solid curves), for $\Gamma=0.1$ and several values of $s=L/l$.
For $s\gg 1$ and $x\ll s$ it simplifies to
\begin{mathletters}
\label{rhoxapprox}
\begin{eqnarray}
&&x=\case{1}{2}{\rm arccosh}\,\tau-\case{1}{2}\Gamma
s(\tau^{2}-1)^{1/2}\cos\sigma,\label{rhoxapproxa}\\
&&\sigma\equiv \pi sN^{-1}\rho(x,s),\;\;\tau\equiv\sigma(\Gamma
s\sin\sigma)^{-1},\label{rhoxapproxb}
\end{eqnarray}
\end{mathletters}%
shown dashed in Fig.\ \ref{rhoplot}. Eq.\ (\ref{rhoxapprox}) has been obtained
independently by Nazarov \cite{Naz93}. For $s=0$ (no disorder), $\rho$ is a
delta function at $x_{0}$, where $\Gamma\equiv 1/\cosh^{2}x_{0}$. On adding
disorder the eigenvalue density rapidly spreads along the $x$-axis (curve a),
such that $\rho\leq N/s$ for $s>0$. The sharp edges of the density profile, so
uncharacteristic for a diffusion profile, reveal the hydrodynamic nature of the
scaling equation (\ref{scaling}). The upper edge is at
\begin{equation}
x_{\rm max}=s+\case{1}{2}\ln(s/\Gamma)+{\cal O}(1).\label{xmax}
\end{equation}
Since $L/x$ has the physical significance of a localization length
\cite{Sto91}, this upper edge corresponds to a minimum localization length
$\xi_{\rm min}=L/x_{\rm max}$ of order $l$. The lower edge at $x_{\rm min}$
propagates from $x_{0}$ to $0$ in a ``time'' $s_{\rm c}=(1-\Gamma)/\Gamma$. For
$1\ll s\leq s_{\rm c}$ one has
\begin{equation}
x_{\rm min}=\case{1}{2}{\rm arccosh}\,(s_{\rm c}/s) -\case{1}{2}[1-(s/s_{\rm
c})^{2}]^{1/2}.\label{xmin}
\end{equation}
It follows that the maximum localization length $\xi_{\rm max}=L/x_{\rm min}$
{\em increases\/} if disorder is added to a tunnel junction. This paradoxical
result, that disorder enhances transmission, becomes intuitively obvious from
the hydrodynamic correspondence, which implies that $\rho(x,s)$ spreads both to
larger {\em and\/} smaller $x$ as the fictitious time $s$ progresses. When
$s=s_{\rm c}$ the diffusion profile hits the boundary at $x=0$ (curve c), so
that $x_{\rm min}=0$. This implies that for $s>s_{\rm c}$ there exist
scattering states (eigenfunctions of $tt^{\dagger}$) which tunnel through the
barrier with near-unit transmission probability, even if $\Gamma\ll 1$. The
number $N_{\rm open}$ of transmission eigenvalues close to one (socalled ``open
channels'' \cite{Imr86}) is of the order of the number of $x_{n}$'s in the
range $0$ to $1$ (since $T_{n}\equiv 1/\cosh^{2}x_{n}$ vanishes exponentially
if $x_{n}>1$). For $s\gg s_{\rm c}$ (curve e) we estimate
\begin{equation}
N_{\rm open}\simeq\rho(0,s)=N(s+\Gamma^{-1})^{-1},\label{Nopen}
\end{equation}
where we have used Eq.\ (\ref{rhoxapprox}).

To test these analytical results we have carried out numerical simulations
similar to those reported in Ref.\ \cite{Mar93}. The disordered normal region
was modeled by a tight-binding Hamiltonian on a square lattice (lattice
constant $a$), with a random impurity potential at each site (uniformly
distributed between $\pm\case{1}{2}U_{\rm D}$). The tunnel barrier was
introduced by assigning a non-random potential energy $U_{\rm B}=2.3\,E_{\rm
F}$ to a single row of sites at one end of the lattice, corresponding to a
mode-averaged transmission probability $\Gamma=0.18$. The Fermi energy was
chosen at $E_{\rm F}=1.5\,u_{0}$ from the band bottom (with
$u_{0}\equiv\hbar^{2}/2ma^{2}$). We chose $U_{\rm D}$ between $0$ and
$1.5\,u_{0}$, corresponding to $s$ between $0$ and $11.7$ \cite{Jon94}. Two
geometries were considered: $L=W=285\,a$ (corresponding to $N=119$), and
$L=285\,a$, $W=75\,a$ (corresponding to $N=31$). In Fig.\ \ref{numplot} we
compare the integrated eigenvalue density $\nu(x,s)\equiv
N^{-1}\int_{0}^{x}dx'\,\rho(x',s)$, which is the quantity following directly
from the simulation. The points are raw data from a single sample.
(Sample-to-sample fluctuations are small, because the $x_{n}$'s are
self-averaging quantities \cite{Sto91}.) The simulation unambiguously
demonstrates the appearance of open channels ($x_{n}$'s near $0$) on adding
disorder to a tunnel barrier, and is in good agreement with our analytical
result (\ref{Eulersol}), {\em without any adjustable parameters}. No
significant geometry dependence was found, as anticipated.

The average resistance of the NS junction is obtained directly from the complex
velocity field $U$ on the imaginary axis. From Eqs.\ (\ref{GNSaverage}) and
(\ref{Udef}) we find
\begin{mathletters}
\label{RNSresult}
\begin{eqnarray}
R_{\rm NS}&=&(h/2Ne^{2}){\textstyle\lim_{\zeta\rightarrow-{\rm
i}\pi/4}}\left(\partial U/\partial\zeta\right)^{-1}\nonumber\\
&=&(h/2Ne^{2})(s+Q^{-1}),\label{RNSresulta}\\
Q&=&\frac{\phi}{s\cos\phi}\left(\frac{\phi} {\Gamma
s\cos\phi}(1+\sin\phi)-1\right),\label{RNSresultb}
\end{eqnarray}
where $\phi\in(0,\pi/2)$ is determined by
\begin{equation}
\phi[1-\case{1}{2}\Gamma(1-\sin\phi)]=\Gamma s\cos\phi. \label{RNSresultc}
\end{equation}
\end{mathletters}%
For $\Gamma\ll 1$ (or $s\gg 1$) Eq.\ (\ref{RNSresultc}) simplifies to
$\phi=\Gamma s\cos\phi$, hence $Q=\Gamma\sin\phi$, in precise agreement with
Ref.\ \cite{Vol93}. The scaling of the resistance with length is plotted in
Fig.\ \ref{GNSplot}. For $\Gamma=1$ the resistance increases monotonically with
$L$. The ballistic limit $L\rightarrow 0$ equals $h/4Ne^{2}$, half the Sharvin
resistance of a normal junction because of Andreev reflection \cite{Blo82}. For
$\Gamma\lesssim 0.5$ a resistance minimum develops, somewhat below
$L=l/\Gamma$. The resistance minimum is associated with the crossover from a
quadratic to a linear dependence of $R_{\rm NS}$ on $1/\Gamma$. The two
asymptotic dependencies are (for $\Gamma\ll 1$ and $s\gg 1$):
\begin{mathletters}
\label{asympt}
\begin{eqnarray}
R_{\rm NS}&=&(h/2Ne^{2})s^{-1}\Gamma^{-2},\;\;{\rm if}\;\; \Gamma s\ll
1,\label{asympta}\\
R_{\rm NS}&=&(h/2Ne^{2})(s+\Gamma^{-1}),\;\;{\rm if}\;\; \Gamma s\gg
1,\label{asymptb}
\end{eqnarray}
\end{mathletters}%
to be contrasted with the classical series resistance \cite{Mar93},
\begin{equation}
R_{\rm NS}^{\rm class}=(h/2Ne^{2})(s+2\Gamma^{-2}), \label{RNSclass}
\end{equation}
which holds if phase coherence is destroyed by a voltage or magnetic field.
Eq.\ (\ref{RNSclass}) would follow from a naive application of Ohm's law to the
NS junction, with the tunnel barrier contributing an additive,
disorder-independent amount to the total resistance. The quadratic dependence
on $1/\Gamma$ in Eqs.\ (\ref{asympta}) and (\ref{RNSclass}) is as expected for
tunneling into a superconductor, being a two-particle process \cite{Hek93}. The
linear dependence on $1/\Gamma$ in Eq.\ (\ref{asymptb}) was first noted in
numerical simulations \cite{Mar93}, as a manifestation of ``reflectionless
tunneling'': It is as if one of the two quasiparticles can tunnel into the
superconductor without reflection. Comparison of Figs.\ \ref{rhoplot} and
\ref{GNSplot} provides the explanation. The resistance minimum occurs when the
lower edge of the density profile reaches $x=0$ (curve c in Fig.\
\ref{rhoplot}), and signals the appearance of scattering states which can
tunnel through the barrier with probability close to one. For $\Gamma s\gg 1$
(curve e), $R_{\rm NS}$ is dominated by the $N_{\rm open}$ transmission
eigenvalues close to one. From Eqs.\ (\ref{GNS}) and (\ref{Nopen}) we estimate
$R_{\rm NS}\simeq h/e^{2}N_{\rm open}=(h/Ne^{2})(s+\Gamma^{-1})$, up to a
numerical prefactor, consistent with the asymptotic result (\ref{asymptb}).

It is essential for the occurrence of a resistance minimum that $1/R_{\rm NS}$
depends {\em non-linearly\/} on the transmission eigenvalues. Indeed, if we
compute the normal-state resistance $1/R_{\rm N}=(2e^{2}/h)\sum_{n}T_{n}$ from
the eigenvalue density (\ref{Eulersol}), we find the linear scaling $R_{\rm
N}=(h/2Ne^{2})(s+\Gamma^{-1})$ for all $\Gamma$ and $s$. The cross-over to a
quadratic dependence on $1/\Gamma$ can not occur in this case, because of the
linear relation between $1/R_{\rm N}$ and $T_{n}$ in the normal state.

In summary, we have presented a scaling theory for the resistance of a
normal--superconductor microbridge. The scaling of the density $\rho$ of
transmission eigenvalues with length $L$ is governed by Euler's equation for
the isobaric flow of a two-dimensional ideal fluid: $L$ corresponds to time and
$\rho$ to the $y$-component of the velocity field on the $x$-axis, with $L/x$
corresponding to the localization length. This hydrodynamic correspondence
provides an explanation for the resistance minimum which is both exact and
intuitive.

We thank Yu.\ V. Nazarov for valuable discussions and R. A. Jalabert for advise
on the numerical simulations. This work was supported by the Dutch Science
Foundation NWO/FOM and by the European Community.

\begin{figure}
\caption[]{
Eigenvalue density $\rho(x,s)$ as a function of $x$ (in units of $s=L/l$) for
$\Gamma=0.1$. Curves a,b,c,d,e are for $s=2,4,9,30,100$, respectively. The
solid curves are from Eq.\ (\protect\ref{Eulersol}), the dashed curves from
Eq.\ (\protect\ref{rhoxapprox}). The resistance minimum is associated with the
collision of the density profile with the boundary at $x=0$, for $s=s_{\rm
c}=(1-\Gamma)/\Gamma$.
\label{rhoplot}}
\end{figure}
\begin{figure}
\caption[]{
Comparison between theory and simulation of the integrated eigenvalue density
$\nu(x,s)\equiv N^{-1}\int_{0}^{x}dx'\,\rho(x',s)$, for $\Gamma=0.18$. The
labels a,b,c indicate, respectively, $s=0,0.7,11.7$. Solid curves are from Eq.\
(\protect\ref{Eulersol}), data points are the $x_{n}$'s from the simulation
plotted in ascending order versus $n/N\equiv\nu$ (filled data points are for a
square geometry, open points for an aspect ratio $L/W=3.8$). The theoretical
curve for $s=0$ is a step function at $x_{0}=1.5$ (not shown). The inset shows
the full range of $x$, the main plot shows only the small-$x$ region, to
demonstrate the disorder-induced opening of channels for tunneling through a
barrier. (Note that, since $T\equiv 1/\cosh^{2}x$, $x$ near zero corresponds to
near-unit transmission.)
\label{numplot}}
\end{figure}
\begin{figure}
\caption[]{
Dependence of the resistance $R_{\rm NS}$ on the length $L$ of the disordered
normal region (shaded in the inset), for different values of the transmittance
$\Gamma$ of the NS interface. Solid curves are computed from Eq.\
(\protect\ref{RNSresult}), for $\Gamma= 1,0.8,0.6,0.4,0.1$ from bottom to top.
For $\Gamma\ll 1$ the dashed curve is approached, in agreement with Ref.\
\protect\cite{Vol93}.
\label{GNSplot}}
\end{figure}


\begin{references}
\bibitem{Jon94} The parameter $s$ was computed numerically by equating $G_{\rm
N}=(2Ne^{2}/h)(1+s)^{-1}$, where $G_{\rm N}$ is the conductance of the
disordered normal region. This is accurate for all $s$, see M. J. M. de Jong,
Phys.\ Rev.\ B (to be published).
\end{references}
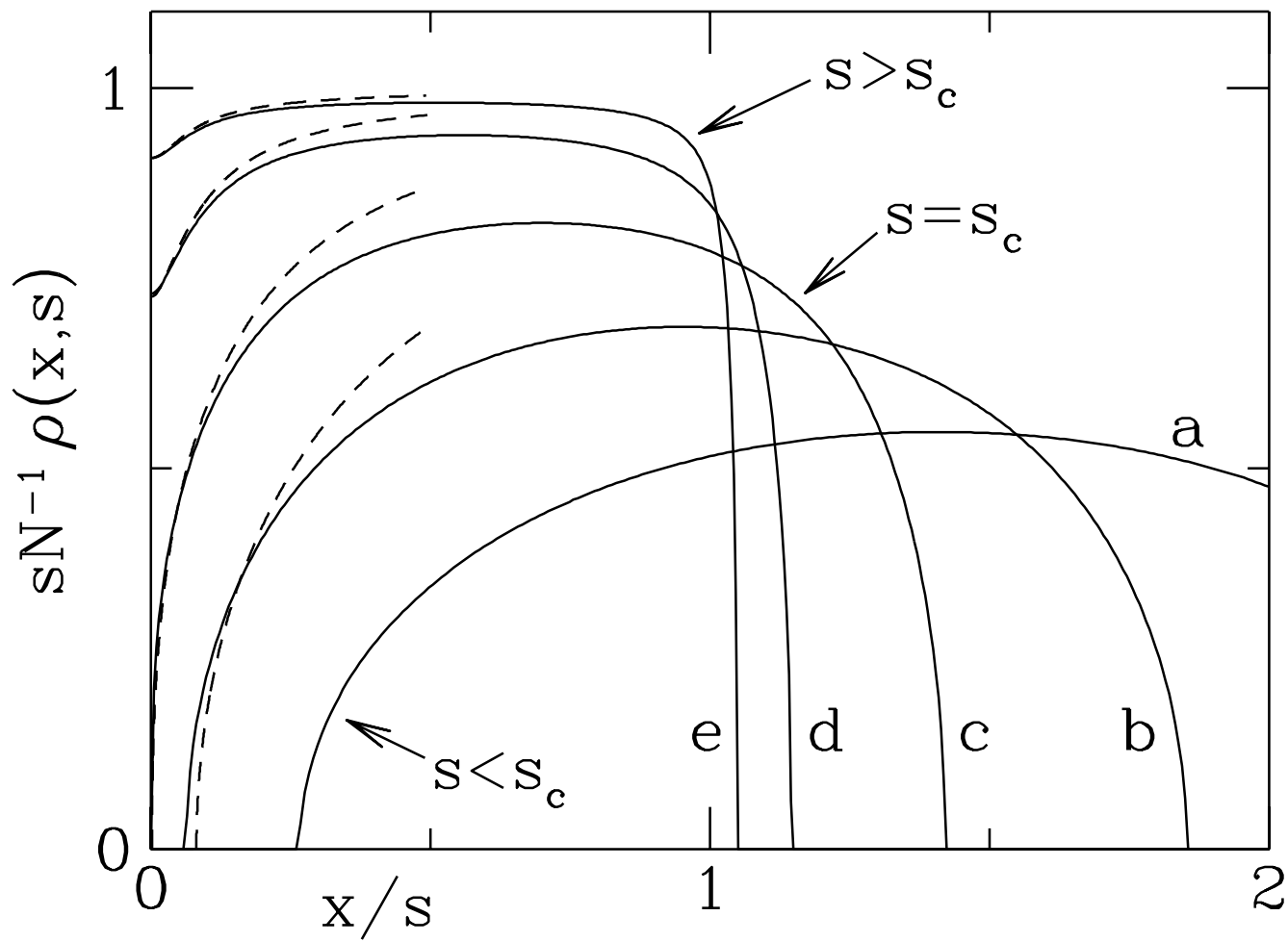
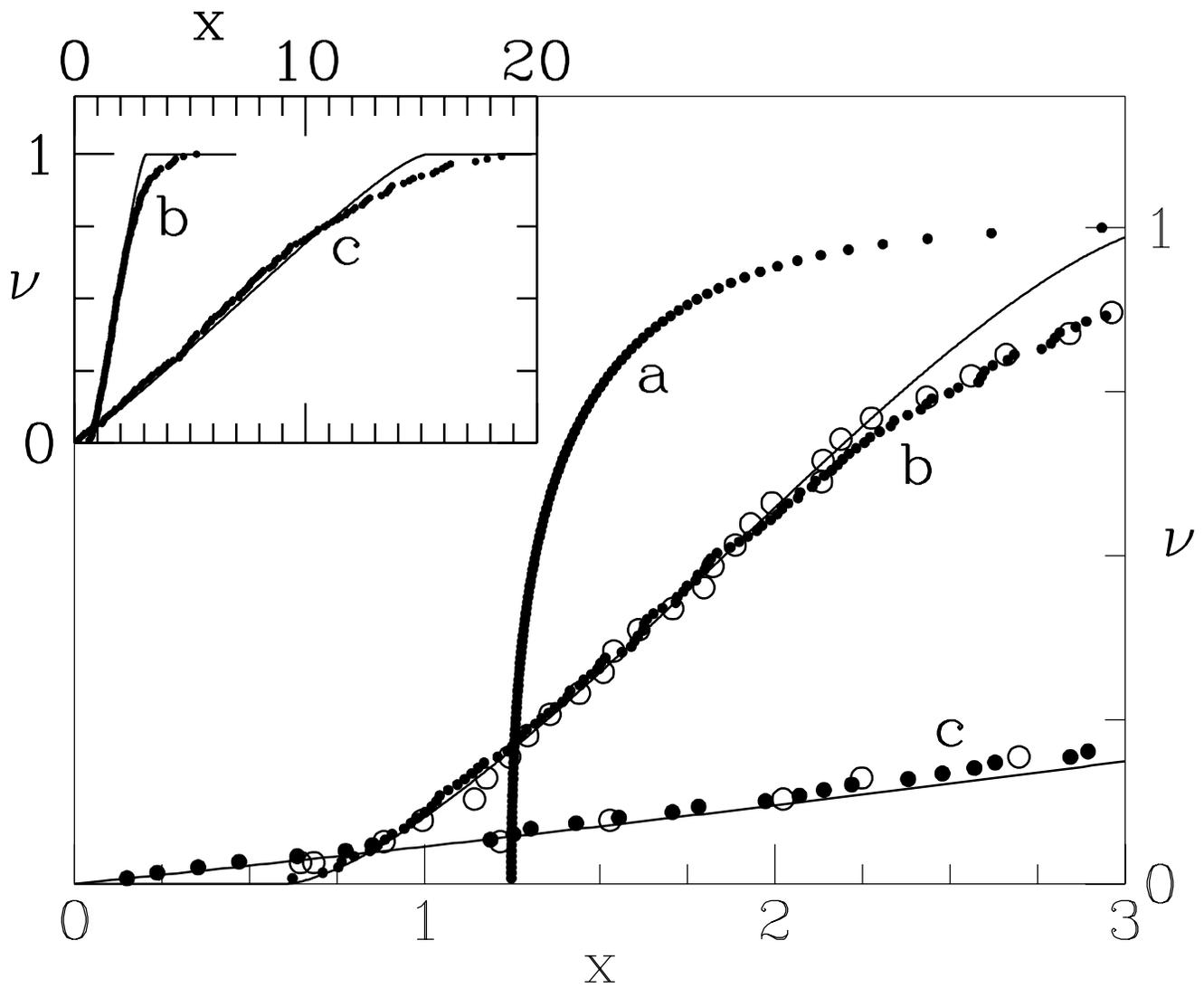
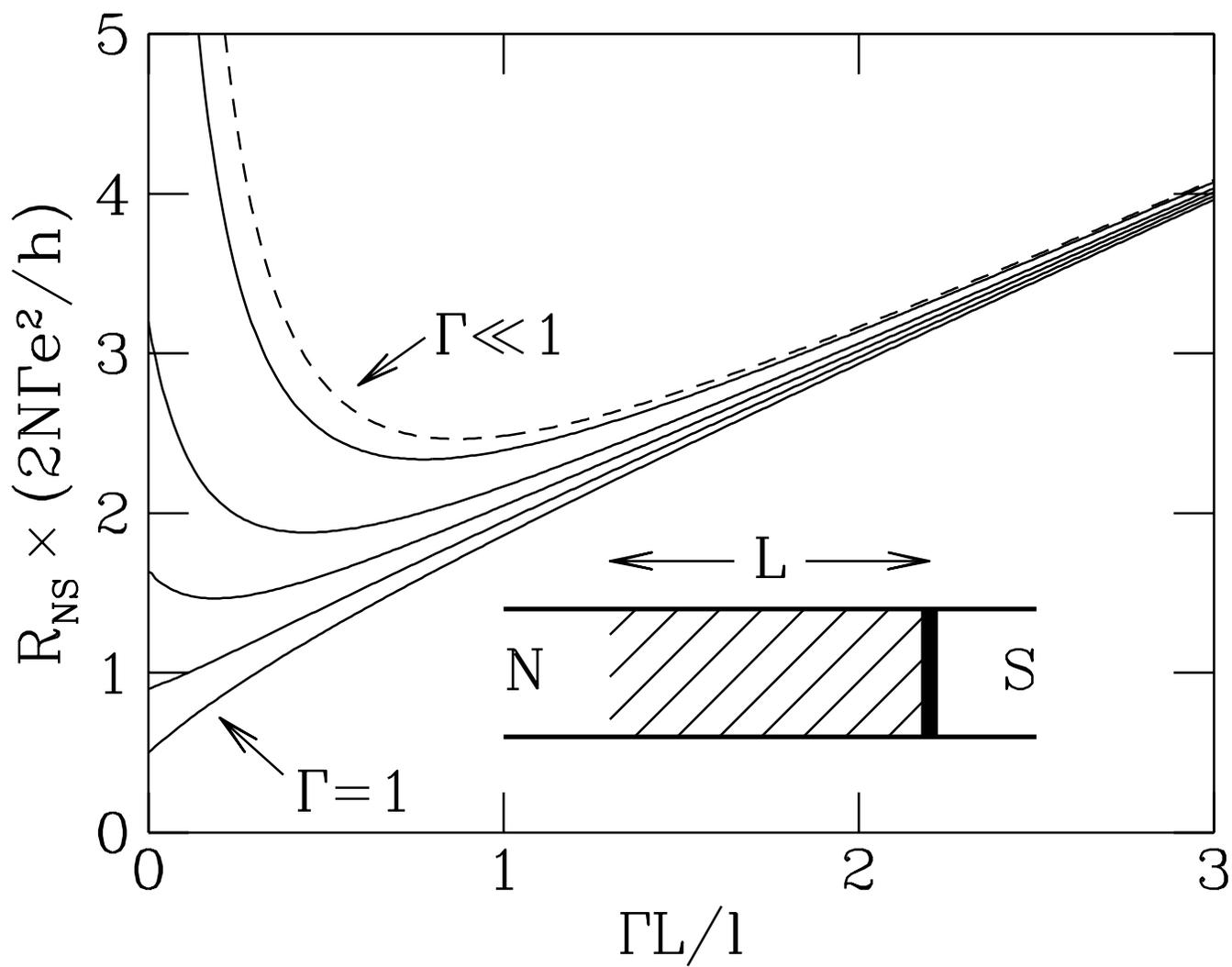
\end{document}